\documentclass[preprint,unsortedaddress]{revtex4-2}
\usepackage{url} 
\usepackage{graphicx}
\usepackage{epstopdf}
\usepackage{subfig,float}
\usepackage{amsmath,amssymb}
\usepackage{natbib}
\usepackage{xcolor}
\usepackage{dcolumn}%
\usepackage{bm}%
\newcommand{\lb}{\label}

\newcommand{\fltr}{\widetilde}
\newcommand{\grad}{{\mbox{\boldmath $\nabla$}}}

\newcommand{\fu}{\fltr\bu}

\newcommand{\btau}{{\mbox{\boldmath $\tau$}}}
\newcommand{\br}{{\bf r}}

\newcommand{\bk}{{\bf k}}

\newcommand{\bu}{{\bf u}}

\newcommand{\bx}{{\bf x}}

\newcommand{\lra}[1]{\langle #1 \rangle }

\def\be{\begin{equation}}
\def\ee{\end{equation}}
\def\lb{\label}
\draft % marks overfull lines with a black rule on the right

\begin{document}

%\preprint{}

\title{Constrained Reversible system for Navier-Stokes Turbulence: evidence for Gallavotti's equivalence conjecture} %Title of paper



\author{Alice Jaccod}
\affiliation{Sorbonne Universit\'e, CNRS, UMR 7190, Institut Jean Le Rond d'Alembert, F-75005 Paris, France}

\author{Sergio Chibbaro}
\affiliation{Sorbonne Universit\'e, CNRS, UMR 7190, Institut Jean Le Rond d'Alembert, F-75005 Paris, France}
%\thanks{}


%\date{\today}

\begin{abstract}
Following the Gallavotti's conjecture, Stationary states of Navier-Stokes fluids are proposed to
   be described equivalently by  alternative
  equations besides the NS equation itself. 
  We propose a model system symmetric under time-reversal based on the Navier-Stokes 
  equations constrained to keep the Enstrophy constant. 
    It is demonstrated through high-resolved numerical experiments that the reversible model evolves to a stationary state which reproduces quite accurately 
    all statistical observables relevant for the physics of turbulence extracted by direct numerical simulations at different Reynolds numbers. 
 The possibility of using reversible models to mimic turbulence dynamics is of practical importance for coarse-grained version of Navier-Stokes equations, as used in Large-eddy simulations.
 Furthermore, the reversible model appears mathematically simpler, since enstrophy is bounded to be constant for every Reynolds. 
  Finally, the theoretically interest in the context of statistical mechanics  is briefly discussed.  \end{abstract}


\maketitle

%\usepackage{geometry}
% \geometry{
% a4paper,
% total={170mm,257mm},
% left=20mm,
% top=10mm,
% right=20mm,
% }
%
%\linespread{1.5}

\paragraph{Introduction}

Non-equilibrium macroscopic systems are generally described in the framework of irreversible Hydrodynamics\cite{lifshitz2013statistical,DeG_84}. 
To explain the arising of an irreversible macroscopic world from the dynamics of the reversible microscopic constituents is one of the central problem of the program of the statistical mechanics~\cite{onsager1931reciprocal,cercignani1988boltzmann,kubo2012statistical}.
In some cases, the Hydrodynamic level is obtained from the microscopic molecular through coarse-graining\cite{kadanoff2000statistical,castiglione2008chaos}, and in this procedure the random effect of small constituents can be replaced by efficient transport coefficients, thanks to the separation of scales.
The laws that thus emerge through the coarse-graining break the fundamental time-reversal symmetry  inherent to the microscopic laws\cite{cercignani1988boltzmann,lebowitz1993boltzmann,chibbaro2014reductionism}.
More generally, the search for a systematic approach remains a key subject of research \cite{spohn2012large,bertini2015macroscopic}

The foremost physical example of irreversible process is given by an incompressible fluid which is described by the Navier-Stokes equations\cite{landau1987course,Fri_95}. In this framework, the molecular effects are represented by viscosity that is also responsible for the dissipation of energy, and may lead to a stationary state when energy is injected.
Most remarkably, even if the limit of vanishing viscosity is taken, the fluid becomes turbulent, strongly chaotic in space and time \cite{Mon_75,Fri_95}, and displays the outstanding feature of ``anomaly dissipation", which means that the mean rate of kinetic energy dissipation $\lra{\epsilon}$ remains finite and independent of $\nu$. Thus, the trace of irreversibility is kept through this singular limit\cite{sreenivasan1997phenomenology,eyink2006onsager}.
The rigorous explanation of such an outstanding feature remains a challenging open issue, and is at the basis of the mathematical problem of the existence and smoothness of the Navier-Stokes in three dimensions \cite{bertozzi2002vorticity,constantin2001some,gallavotti2013foundations}.

In the last decades, a renewed interest has been put on this fundamental question also using experiments and numerical simulations. Notably, non-trivial features of irreversibility have been found in Lagrangian statistics \cite{xu2014flight}, and such extreme events have been unveiled that they have been related to possible singularities in Navier-Stokes equations \cite{saw2016experimental,dubrulle2019beyond}.
This research investigates from different angles the possibility for gradients to become unbound in the limit of vanishing viscosity.
A big problem of such an approach is the asymptotic nature of turbulence, which make difficult to disentangle in actual experiments 
Reynolds-number effects from genuine features\cite{iyer2018steep,iyer2019circulation}.
An alternative approach was proposed some years ago by Gallavotti through the conjecture that the same system can be described by different yet equivalent models,
notably for fluids\cite{gallavotti1997dynamical}. 
In particular, phenomenological irreversible macroscopic systems could be described by suitable reversible models, at least in some respect.
This idea was rooted in several developments in statistical physics. For instance, Molecular Dynamics simulations have shown that for large number of particles (“thermodynamic limit”), most statistical properties of the “test” system do not depend on 
the details of the different models used through the application of thermostats \cite{evans2008statistical,hoover2012computational}.
Most importantly, these results have emphasised the difference between reversibility and dissipation, since several thermostats are reversible, while the dynamics being dissipative.

The possibility to use a time-reversible model to obtain turbulent features was pioneered in~\cite{she1993constrained}, and then conjectured in a more formal way by Gallavotti~\cite{gallavotti2014nonequilibrium,gallavotti2019nonequilibrium}.
This conjecture has been called of equivalence of dynamical ensemble, to clearly point the analogy with ensembles in equilibrium statistical mechanics\cite{gallavotti1995dynamical}.
In this framework, while energy is fixed in microcanonical ensemble, it fluctuates in the canonical one but depends on the fixed temperature. If the temperature is equal to the average kinetic energy in the microcanonical ensemble, the statistics of most observables are     
equal or close. More strongly, in the thermodynamic limit, $N\rightarrow \infty$ with $\rho=N/V=const$, any local observable, that is related to a finite region of the phase space, is equal in the two ensembles.
Following this picture, it has been proposed to replace the constant viscosity with a fluctuating one that would make possible to have a new global invariant  for the system.
The thermodynamic limit is obtained in the case $1/\nu \rightarrow \infty$.
%, where $Re=u L /\nu $ is the Reynolds number dimensionless  
Since in this fully turbulent limit, the system is highly chaotic and exhibits a random behaviour, it may be plausible to conjecture that it may be described by an invariant distribution, as already postulated by Kolmogorov in his founding works\cite{kolmogorov1941local,kolmogorov1941equations,kolmogorov1941dissipation}.

The conjecture has been directly tested in small 2D systems\cite{gallavotti2004lyapunov,gallavotti2019nonequilibrium},  for the Lorenz model \cite{gallavotti2014equivalence}, in shell models\cite{biferale1998time,biferale2018equivalence}. Recently, a model obtained by imposing the constraint that turbulent kinetic energy is conserved has been analysed  in 3D turbulence with a small number of modes\cite{shukla2019phase}. 
Parallel tentatives have been made to test the consequences, namely the fluctuation relations in different systems\cite{ciliberto1998experimental,shang2005test,bandi2009probability,zonta2016entropy}.
%Plausibility of the conjecture is given also by important recent studies which show that the dynamics at large scales is basically insensitive to the dissipation at small-scales \cite{dallas2015statistical,alexakis2019thermal}.
While these studies provide motivation to the present work and have given important insights, a clear demonstration of the validity of the Gallavotti's conjecture still lacks.

Different equivalent models may in principle be proposed \cite{gallavotti2019nonequilibrium}, yet considering the physics of Turbulence the reversible model should be related to the dissipation anomaly, where the average rate of dissipation is defined as $\lra{\epsilon} \equiv \lra{\nu \vert\Delta u \vert ^2} = 2 \nu \Omega$, where $\Omega=\lra{\mathbf{\omega}^2}$ is the enstrophy, expressed in terms of the vorticity $\mathbf{\omega} = \nabla \times \mathbf{u}$~\cite{Fri_95}.
In analogy with statistical mechanics \cite{huang1963statistical}, 
we consider the irreversible distribution as the canonical ensemble with $\nu$ corresponding to $\beta=(k_BT)^{-1}$, and therefore we build the analogous to the microcanonical ensemble taking the enstrophy $\Omega$ as fixed. corresponding to the energy, and letting  $\nu$ fluctuating.
If the equivalence holds, the average rate of dissipation should be the same.

The purpose of the present work is to show to which extent the Gallavotti conjecture is accurate, using high-resolution numerical experiments at different Reynolds numbers.

To find that the Gallavotti's  conjecture holds has importance from different point of views.
From a mathematical point of view the conjecture  is related to the issue of a rigorous proof of existence of unique  solutions of the Navier-Stokes equations \cite{constantin1988navier,temam2001navier,gallavotti2013foundations}. 
Indeed, the reversible model proposed should admit a smooth solution, since the vorticity remains bounded for any value of the viscosity. While the original mathematical problem would remain open, the conjecture should provide an answer at least from the statistical point of view, since the same statistical results can be obtained with a well-posed set of equations.
From the physical point of view, this conjecture arose in relation to the development of a general framework for non-equilibrium problems in statistical mechanics\cite{evans1993probability,gallavotti1995dynamical-a,marconi2008fluctuation}, formally based on the chaotic hypothesis~\cite{ruelle1995turbulence,gallavotti1996extension}, which was meant conceptually to be a founding hypothesis for non-equilibrium dissipative systems analogous to the Ergodic one for equilibrium non-dissipative ones~\cite{gallavotti1995dynamical,lebowitz1999gallavotti}.
The main difficulty is that the general theory applies only to time-reversible dynamical systems, whereas NS is not. However if the conjecture holds true, it  means that many non-equilibrium systems, and most notably turbulent fluids could be considered \emph{in practice} as reversible as far as statistical observables are considered, and therefore Gallavotti-Cohen theory could be applied to the correct observables.
%discovered that under suitable assumptions these SNSsatisfy a certain symmetry, which they doobed a “fluctuationtheorem”. Assum-ing that the dynamics satisfiestime reversal invarianceand is sufficiently chaotic,so that the SNS is given by an SRB measure, they prove that the probability dis-tribution for the phase space contraction averaged along a trajectory over thetime spanτhas, for largeτ, a highly non-obvious symmetry, whose specific formwill be given below. Near equilibrium the fluctuation theorem implies Onsagerreciprocity and the Einstein relations. 
From the applicative point of view, multi-scale approach is crucial to tackle  complex systems with reduced models, like in climate and meteorological sciences.
In this case, only large-scales can be simulated and small-scales are modelled often in an irreversible dissipative way\cite{lesieur1996new,sagaut2006large}.
The present study aims to give some insights on new possible way to propose reversible models, since it is known that such models may better describe the cascade process \cite{meneveau2000scale}.

%In this work, we provide a systematic numerical study of the conjecture using highly-resolved numerical simulations up to $1024^3$ points for three different values of the viscosity. 
%In all cases, we have considered the physical relevant situation in which cascade takes place.
%which gives a clear demonstration of the validity of the conjecture.

\paragraph{Mathematical Framework}
Let us consider a dynamical  system:
\begin{equation}
\dot{\mathbf{x}} = h(\mathbf{x}) - \nu L \mathbf{x} + f(\mathbf{x})
\label{irr}
\end{equation}
with $h(\mathbf{x}) = h(-\mathbf{x})$. If $\nu = 0$ and $f$ is even, the equation has a symmetry of temporal inversion $I$ if the solution operator $ \mathbf{x}\rightarrow S_t \mathbf{x}, \mathbf{x} \in \mathcal{R}^N$ and the map $I$  are such that $I^2=I, S_tI = IS_{-t}$. In the case $\nu > 0$, assuming that $|\mathbf{x} \cdot h(\mathbf{x})| \le \Gamma (\mathbf{x} \cdot L\mathbf{x})$, the motion is asymptotically confined,  i.e $(\mathbf{x} \cdot L\mathbf{x}) \le \frac{G}{\nu}$ and a stationary state will exist, described by an invariant distribution of probability $\mu_{\nu}$\cite{ruelle1989chaotic,ruelle1995turbulence}, %called \textit{viscosity ensemble }. 
For each $\nu$, the distribution defines a   \emph{ nonequilibrium ensemble} $\mathcal{E}_\nu$.
% whose the various elements are parametrized by a different value of $\nu$.\\
Now, let us consider a new equation 
in which the viscous coefficient in Eq. (\ref{irr}) is replaced by a multiplier such that a relevant observable $\mathcal{O}$
is maintained as a constant of motion.
As explained shortly, the relevant observable for NS is the following $\mathcal{O}(\mathbf{x})=(\mathbf{x}\cdot L\mathbf{x})$,
and the equation becomes
$ \dot{\mathbf{x}} = h(\mathbf{x}) - \alpha L\mathbf{x} + f(\mathbf{x})$, with
\begin{equation}
\alpha = \frac{Lx \cdot h(x)}{Lx \cdot Lx}~.
\end{equation}
This equation is reversible under time-reversal.
Its stationary states form a collection of new \textit{reversible viscosity ensemble} $ \mu_{{\mathcal{O}}}$ labelled by the value of the constant of motion. 
Denoting $\lra{}_\nu, \lra{}_\mathcal{O}$ the averages over the two distributions,
the content of the Gallavotti's Conjecture is the following: for small enough $\nu$, it can be expected that the system is highly chaotic and $\alpha(x)$ fluctuates wildly leading to a multi-scale or homogenisation phenomenon\cite{sanchez1980non,castiglione2008chaos}, that is a large class of observables have the same statistics in the two distributions 
$\mu_\nu$ and $\mu_\mathcal{O}$ provided that $\lra{\alpha}_\mathcal{O}=\nu$ or equivalently $\lra{\mathcal{O}}_\nu=\textrm{O}$.
This means that for a set of macroscopic observables $\Theta$ the averages are the same for the reversible and irreversible dynamics, such that $\lra{\Theta}_{\mathcal{O}}=\lra{\Theta}_{\nu}$, in the limit of vanishing viscosity $\nu\rightarrow 0$.
This proposal is called  \textit{conjecture of equivalence} and assures the statistical equivalence in the frame of dynamical systems between irreversible and reversible formulation.

\paragraph{Reversible Hydrodynamics}
We consider here an incompressible fluid, with constant density $\rho = 1$, subjected to viscosity and an external forcing term. The motion is described by the NS equation:
\begin{equation}
\partial_t \boldsymbol{u} + (\boldsymbol{u} \cdot \nabla) \boldsymbol{u} = - \nabla p + \nu \nabla^2\boldsymbol{u} + \boldsymbol{f} \qquad\qquad \nabla\cdot \boldsymbol{u} = 0
\label{ns}
\end{equation}
where $\nu$ is the cinematic viscosity, $p$ the pression and $\boldsymbol{f}$ a forcing term which acts at large scales. Clearly, the dissipative term breaks up the symmetry for temporal inversion, i.e the equation is not invariant under the transformation:
$\mathcal{T}: t \rightarrow -t; \boldsymbol{u} \rightarrow \boldsymbol{-u}.
$
%The standard irreversible NS model will be labeled as INS.
As explained above, the corresponding reversible model is obtained replacing the viscosity coefficient  $\nu$ with a time-dependent term which makes the equation invariant under the symmetry $\mathcal{T}$. 
Imposing the conservation of enstrophy $\Omega \equiv \int_{\mathcal{V}}|\nabla\times\boldsymbol{u}|^2  d\boldsymbol{x}$, the equation (\ref{ns}) becomes the reversible Navier-Stokes (RNS)  $\partial_t \boldsymbol{u} + (\boldsymbol{u} \cdot \nabla) \boldsymbol{u} = - \nabla p +  \alpha[\boldsymbol{u}]\nabla^2\boldsymbol{u} + \boldsymbol{f}$ 
with the \emph{fluctuating} viscosity defined as
\begin{equation}
\alpha[\boldsymbol{u}] = \frac{\int_{\mathcal{V}}^{}[\boldsymbol{g}\cdot \omega + \omega \cdot (\omega \cdot \nabla)\boldsymbol{u}] \,d\boldsymbol{x}}{\int_{\mathcal{V}}^{}(\nabla\times\mathbf{\omega})^2\, d\boldsymbol{x}}
\label{ao}
\end{equation}\\
where the vorticity $\mathbf{\omega}=\nabla\times\boldsymbol{u}$, and $\boldsymbol{g} = \nabla\times \boldsymbol{f}$ are used.
In such a case, we have that the enstrophy
 {is a constant of motion}.
Some details more about the theory are given in the supplemental material.
%It is remarkable to underline that the conjecture of equivalence is strictly linked with the \textit{chaotic hypothesis}, formulated in (\cite{cohen}): in the limit of large Reynolds numbers, the system is chaotic and $\alpha(\boldsymbol{u})$ is a \textit{self-averaging} quantity that tends to a constant value $\nu = 1/R$. Moreover, the fact that $\alpha(\boldsymbol{u}) = -\alpha(\boldsymbol{-u})$ assures that the invariance for temporal inversion is rebuild. The so called limit $Re \rightarrow \infty$ is equivalent to the thermodynamical limit at which $N \rightarrow \infty$.

\paragraph{Numerical demonstration}
We perform numerical simulations of the 3D NS and the 3D RNS Eqs. by using the  efficient, parallel finite-volume numerical code Basilisk\footnote{{http://basilisk.fr}}.
The velocity field $\mathbf{u}$ is solved inside a cubic domain of side $2 \pi$, and is prescribed to be triply-periodic. 
The temporal dynamics is integrated via a third-order Adam-Bashfort method.
Both the RNS and NS runs are initiated from the  Taylor-Green velocity field\cite{brachet1983small}.
In order to obtain statistically steady states, we inject energy in the system by using the Taylor-Green forcing\cite{brachet1984taylor}. 
%More details are given in the supplementary material. 
As usual in isotropic turbulence, we characterise the  flow by using the dimensionless Reynolds number based on the Taylor length \cite{Fri_95} $R_\lambda=u_{rms} \lambda /\nu$. We have performed three simulations at $R_\lambda=30,100,300$.
All simulations are carried out so that the smallest scale $\eta$ is very well resolved ($\Delta x/\eta \lesssim 1$ in all cases), and the corresponding number of points  used are $N=256,512,1024$.
%----------------------------------------------------------------------------
%----------------------------------------------------------------------------
\begin{figure*}
\centering
\includegraphics[width=\textwidth]{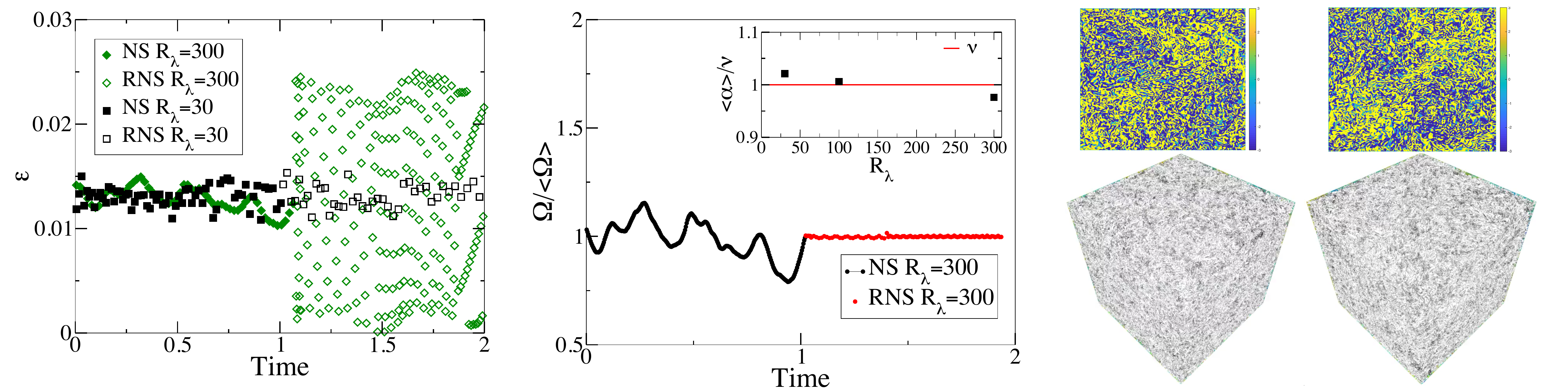}
\caption{Time-dynamics of some observables in the irreversible NS and then switched in the Reversible model. 
(a) Comparison between the time evolution of dissipation rate $\epsilon$ in the irreversible NS and the reversible RNS model for different Reynolds numbers. Time is normalised with the large-scale  (integral) characteristic time-scale. The case at $R_\lambda=100$ is very similar to the $R_\lambda=30$ one and is not shown for the sake of clarity.
(b) Time-dynamics of Enstrophy $\Omega$ normalized by its average value at the highest $R_\lambda$. In the reversible model the enstrophy is kept constant. 
In the inset, The mean value of the fluctuating viscosity in the reversible model normalised by the constant viscosity value is show at different Reynolds numbers.
(c) Visualisation of the vorticity field for the NS (left panel) and RNS right panel. The 3D are obtained with the $\lambda_2$ criterion. The snapshots are the vorticity field at a given time at the centre of the cube.
%Notice that $ \alpha$ is not positive definite: The occurrence of negative values, highlighted with a thick red line, corresponds to instances in which the dissipative terms inject energy into the system.
}
\label{fig1}
\end{figure*}
%\begin{figure}[h]
%\includegraphics[width=0.45\textwidth]{fig1a}
%\includegraphics[width=0.45\textwidth]{fig1b}
%\includegraphics[width=0.45\textwidth]{fig1c}
%\caption{Time-dynamics of some observables in the irreversible NS and then switched in the Reversible model. 
%(a) Comparison between the time evolution of dissipation rate $\epsilon$ in the irreversible NS and the reversible RNS model for different Reynolds numbers. Time is normalised with the large-scale  (integral) characteristic time-scale. The case at $R_\lambda=100$ is very similar to the $R_\lambda=30$ one and is not shown for the sake of clarity.
%(b) Time-dynamics of Enstrophy $\Omega$ normalized by its average value at the highest $R_\lambda$. In the reversible model the entrophy is kept constant. 
%In the inset, The mean value of the fluctuating viscosity in the reversible model normalised by the constant viscosity value is show at different Reynolds numbers.
%%Notice that $ \alpha$ is not positive definite: The occurrence of negative values, highlighted with a thick red line, corresponds to instances in which the dissipative terms inject energy into the system.
%}
%\label{fig1}
%\end{figure}
In figure \ref{fig1} the phenomenology of both models is illustrated by displaying the dynamics of the dissipation-rate and of the Enstrophy  at different Reynolds numbers.
It is seen from Fig. \ref{fig1}a that the reversible model at high Reynolds numbers shows wild fluctuations in $\varepsilon=2 \alpha \Omega$ because of the behaviour of the fluctuating viscosity $\alpha$. At more moderate Reynolds the behaviour is practically indistinguishable between NS and RNS.
It is worth noting some sporadic negative events in dissipation at high Reynolds, meaning that there is sometime injection of energy by viscosity.
The first prediction of the conjecture is the reciprocity property which states that if enstrophy is taken fixed $\Omega_{RNS}=\langle \Omega \rangle_{NS}$, then $\nu=\lra{\alpha}$. This is a prerequisite for the conjecture of equivalence.
 In Fig. \ref{fig1}b it is shown that this is true within the numerical errors (about $1\%$) at all Reynolds.
 From a more qualitative point of view, Fig 1c shows that also the geometrical features of the turbulent flow are practically indistinguishable in the reversible and irreversible dynamics.
%Interestingly, the phenomenology is similar to what found in shell models \cite{}.
%----------------------------------------------------------------------------
%----------------------------------------------------------------------------
%----------------------------------------------------------------------------

\begin{figure}[h]
\includegraphics[width=0.45\textwidth]{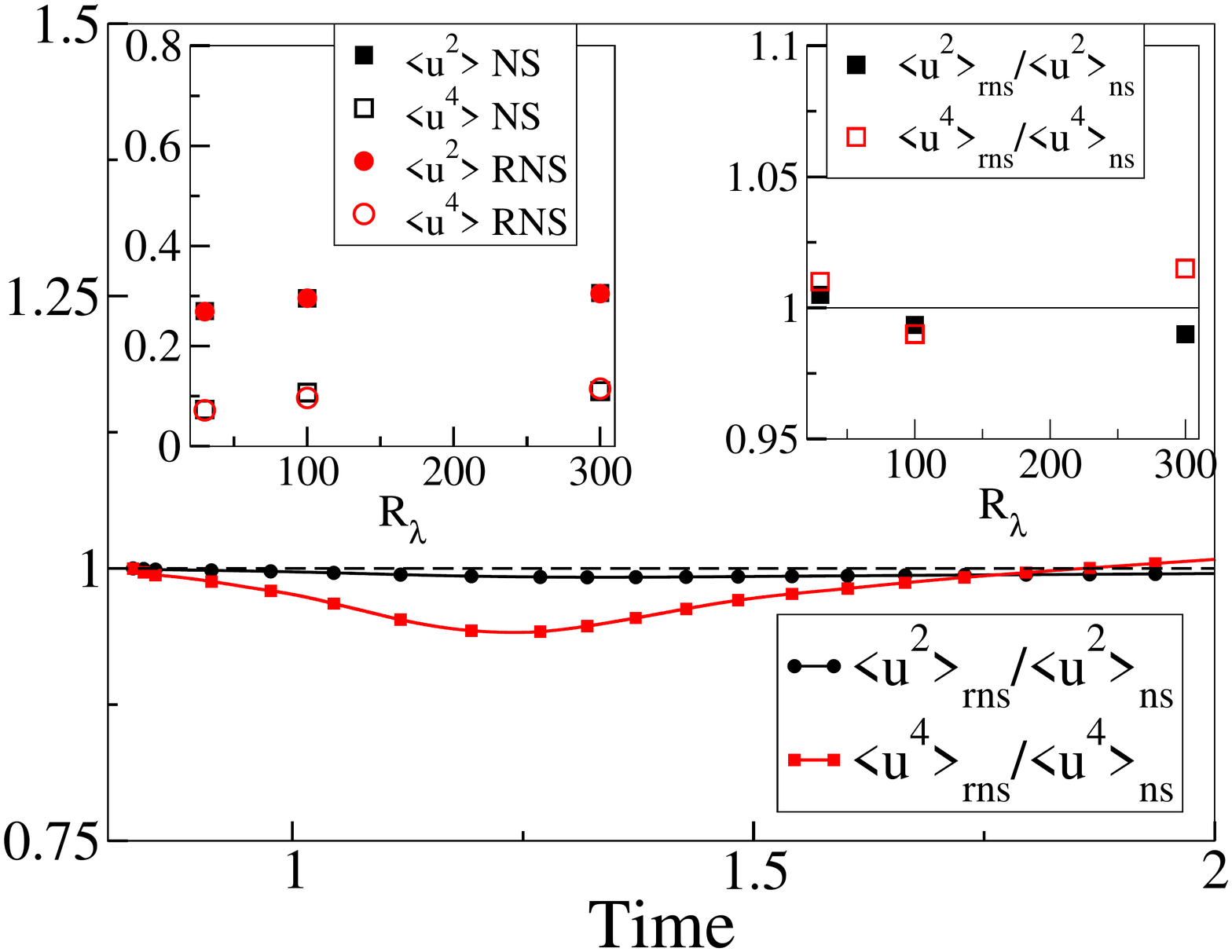}
\includegraphics[width=0.45\textwidth]{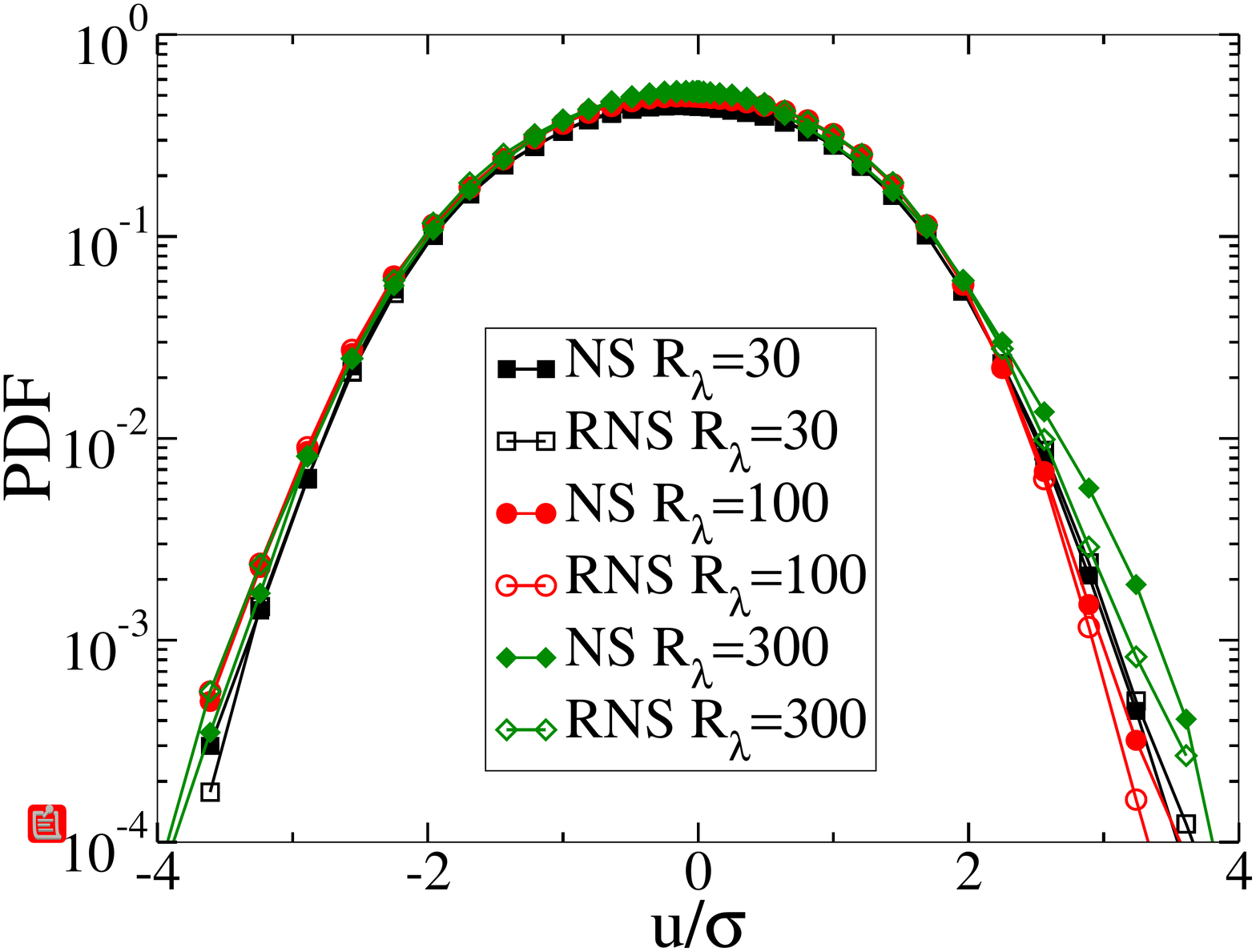}
\caption{Test of equivalence:
(a): running average of the ratio between the second statistical moment and fourth moment of the reversible model with respect to the irreversible one at $R_\lambda=300$. In the inset, comparison of the same moments for NS 
(closed symbols) and RNS (open symbols) as a function of the Reynolds number.
Right inset: Same moments of a velocity field component pertaining to the large scales, only the $k=3$ mode of the Fourier transform of the field is taken.
(b) One-point pdf of the velocity field
% Gaussian curve is displayed for reference.
 at different Reynolds number. }
\label{fig2}
\end{figure}
The stringent test of the conjecture is about the equivalence of statistical properties of local observables (where locality is intended in momentum space).
Since dissipation takes place at small scales, the observables are local if they reside at large scale only. 
We compare in Fig. \ref{fig2}a the second and fourth statistical moment of the the velocity field.
We have computed them both from the whole field, that is containing all the wave-modes, 
and for a field pertaining only the large scale where only the third Fourier mode is taken.
 While the instantaneous value wildly oscillate, the mean values converge rapidly to the irreversible value.
The conjecture focuses in principle on the \emph{local} (large-scale) observables, but remarkably we find 
that also the global statistical moments converge to the irreversible value within the numerical errors.
To further corroborate this property, we have studied the entire one-point pdf of the velocity, see Fig. \ref{fig2}b, which indicates that the irreversible and reversible PDF of the velocity field are in quite good agreement, with only some minor discrepancy in right tail for the highest $R_\lambda$, which are yet of the order of the statistical error.
%These results demonstrate the validity of the conjecture of equivalence and suggest that the conjecture can be extended to less local variables.

%----------------------------------------------------------------------------
\begin{figure}[h]
\includegraphics[width=0.45\textwidth]{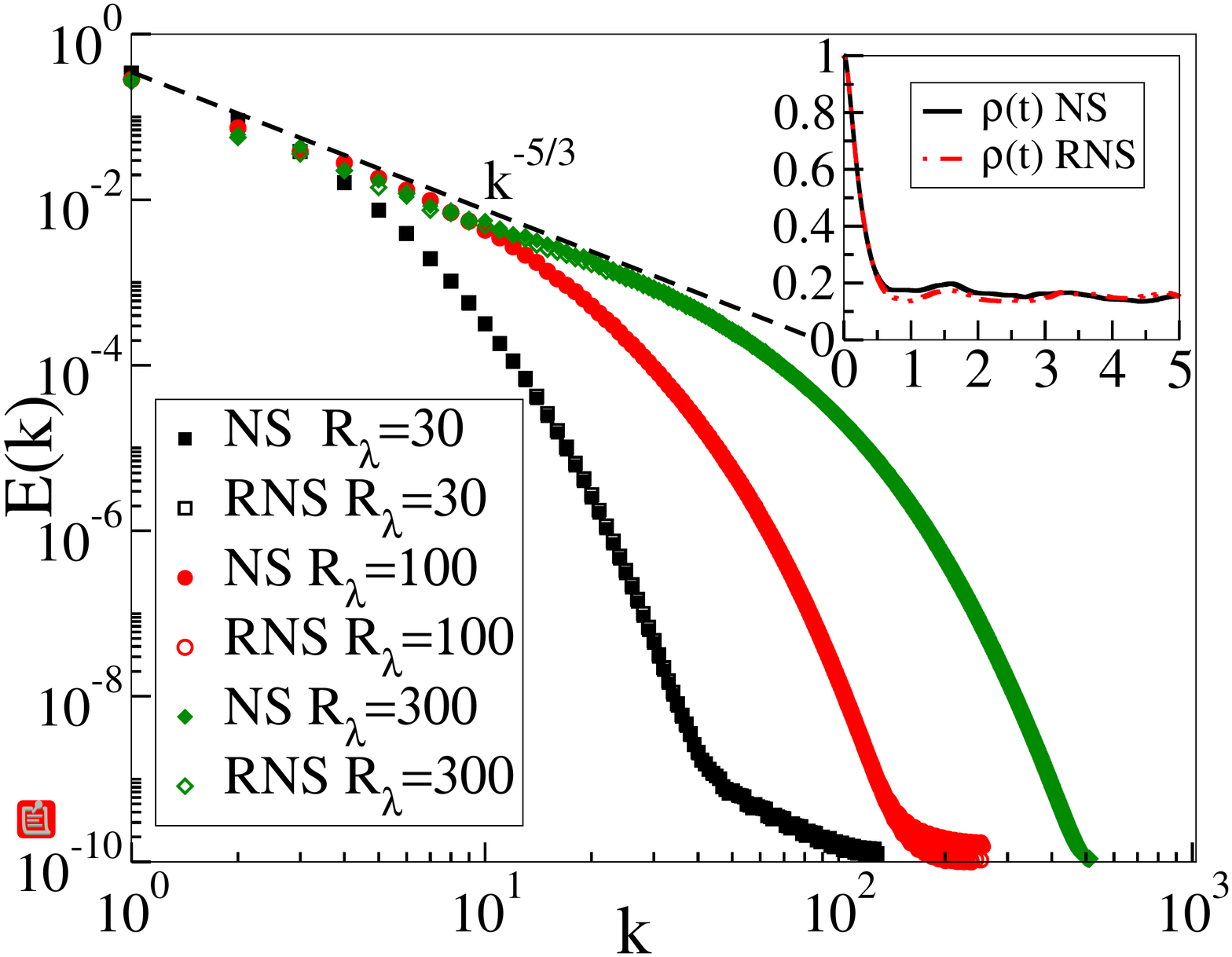}
\includegraphics[width=0.35\textwidth,angle=90]{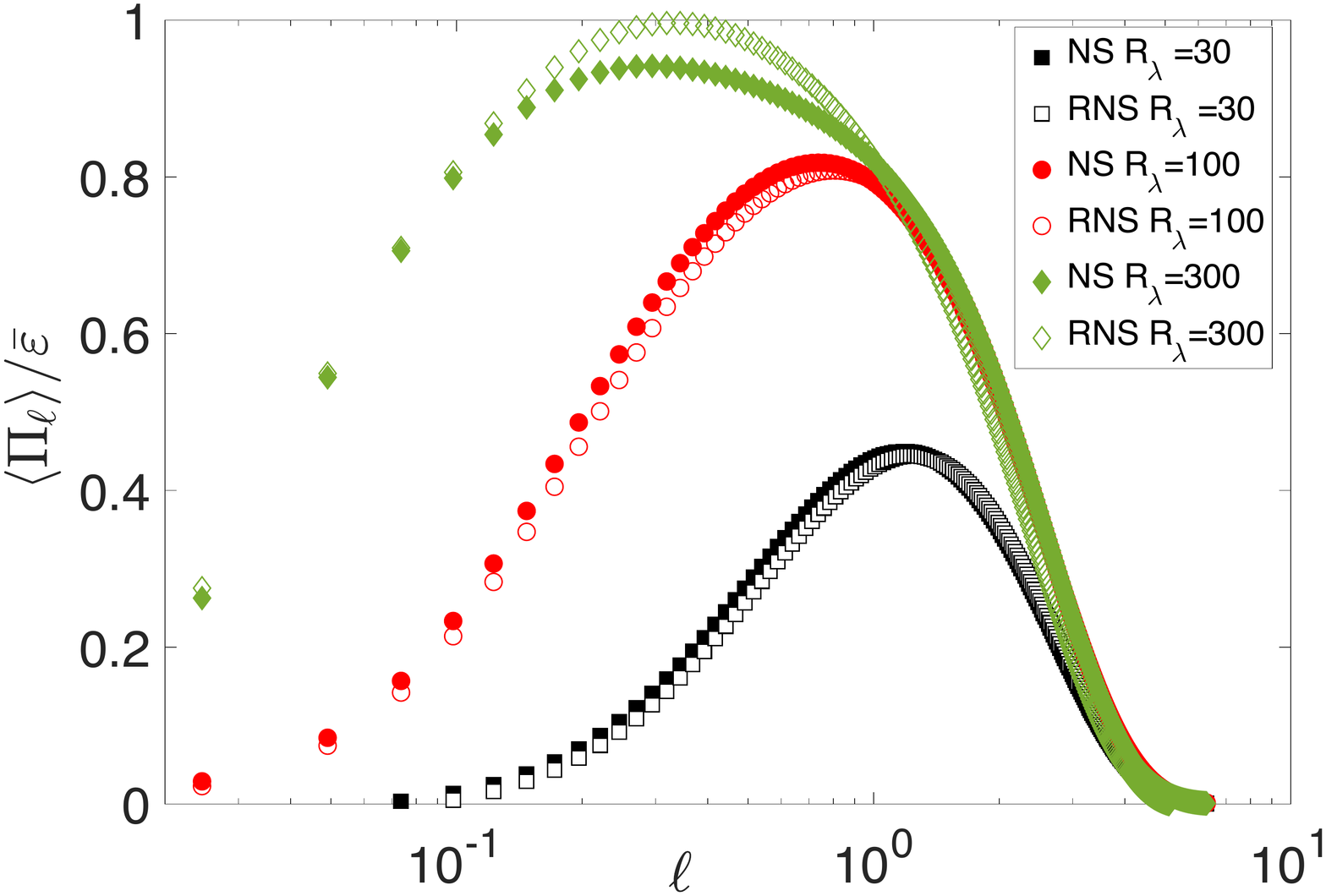}
\caption{(a) Time-average of the energy spectrum $E(k,t) \equiv 1/2 \sum_{\b k} \vert {\bf u}({\b k},t)\vert^2$,
at different Reynolds number for the irreversible and reversible models.  
In the inset the normalised auto-correlation in time of the velocity for both NS and RNS $\rho(t)=\langle u(t_0) u(t_0+t)\rangle/\sigma_u$, at $R_\lambda=300$. (b) Scale-by-scale flux of energy normalised by the mean dissipation-rate.}
\label{fig3}
\end{figure}
%----------------------------------------------------------------------------
Key for the dynamic of turbulence are the two-point statistical observables~\cite{Mon_75,kraichnan1971inertial,Fri_95}.
We show both velocity time-correlation and one-dimensional Energy spectrum in Fig. \ref{fig3}a.
An excellent agreement between irreversible and reversible models is found at all scales.

Even more important is the scale-by-scale flux of energy, which describes the cascade of energy.
We compute the scale-by-scale flux  from the coarse-graining of the Navier-Stokes equation (\ref{ns}) as \cite{germano1992turbulence,eyink2006onsager} 
\begin{equation}
\Pi_\ell({\mathbf x}) \equiv -(\frac{ \partial \overline{u}_{i}}{\partial x_j})\tau_{ij}~,~~\text{with}~~({\boldmath \tau}_\ell)_{i,j} = 
\overline{(u_i u_j)}_\ell -
(\overline{u}_\ell)_i (\overline{u}_\ell)_j~,
\end{equation}
where the dynamic velocity field $\bu$ is spatially (low-pass) filtered over a scale $\ell$ to obtained a filtered value:
$\fltr{\bu}_\ell(\bx) = \int d^3 r\, G_\ell(\br) \bu(\bx+\br)$
where $G_\ell$ is a smooth filtering function, spatially localized and such that $G_\ell (\vec r) = \ell^{-3}G(\vec r/\ell)$ and $G$ satisfies
$\int d\vec r \ G(\vec r)=1$, and $\int d\vec r \ \vert \vec r \vert ^2 G(\vec r) = \mathcal{O}(1)$.
The results of the flux for the different numerical experiments are displayed in Fig \ref{fig3}b up to scale $\ell=2\pi/256$. 
The global behaviour is the same as obtained in analogous pseudo-spectral simulations \cite{chen2003joint,alexakis2020local}, but what is important is that the fluxes of the reversible and irreversible model are the same at all scales, and at all $R_\lambda$.
A small discrepancy is present at $R_\lambda=300$ in the inertial range, which is probably due to different statistical convergence.
These results show unambiguously that the mechanics of turbulence is the same with both irreversible and reversible model.
%----------------------------------------------------------------------------

\begin{figure}[h]
\includegraphics[width=0.5\textwidth]{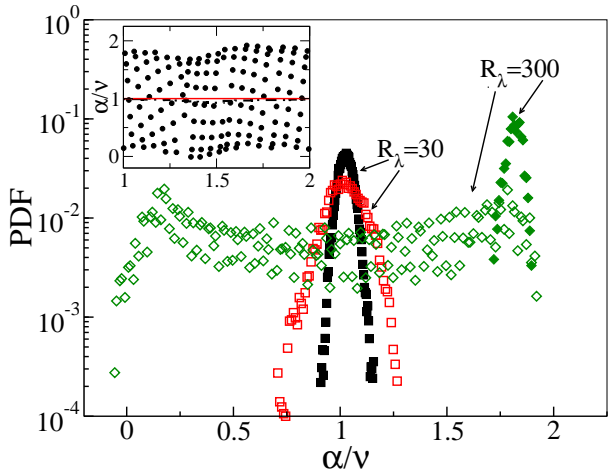}
\caption{Probability density function of $\alpha$. As in the previous figures, filled symbols are for NS and  void symbols for RNS.
The insets on the left show the corresponding typical time evolutions of $\alpha$ at $R_\lambda=300$. }
\label{fig4}
\end{figure}
%----------------------------------------------------------------------------
Finally, we analyse the statistics of the time-fluctuating viscosity $\alpha$, shown if Fig. \ref{fig4}. 
With respect to the equivalence conjecture, the sole crucial feature is that $\lra{\alpha}=\nu$, as shown in Fig. \ref{fig1}. 
The statistics of  $\alpha$ are interesting per se in connection with the symmetry of fluctuations given by the Fluctuation relations for time-reversible dynamical systems\cite{marconi2008fluctuation,gallavotti2014nonequilibrium}. Indeed, $\alpha$ is the most important quantity from the statistical mechanics angle, since it is related to the entropy production in the time-reversible model \cite{gallavotti2019nonequilibrium}.
We plot the PDF of $\alpha$ computed using formula (\ref{ao}) during the reversible dynamics as well as that computed in the irreversible one at different Reynolds number. 
In the reversible dynamics, $\alpha$ fluctuates around the ``canonical" value $\nu$,
and the the variance increases with the Reynolds number.
At low and moderate Reynolds numbers no negative event is recorded. Instead some are found at $R_\lambda=300$, when distribution turns out to be much more flatter.  
As discussed in recent works \cite{biferale2018equivalence,shukla2019phase},  the limit $R_\lambda\rightarrow\infty$ and $N\rightarrow\infty$ is singular and the different behaviour of the PDF reflects that.
Furthermore, our results show that in the cascade regime analysed here, it is difficult  to observe extreme events on a \emph{reasonable} observation-time, notably at small $R_\lambda$.
As expected for the 3D case \cite{gallavotti1997dynamical}, the statistics of $\alpha$ of the reversible and irreversible dynamics are qualitatively different.
The entropy production should be the same in both dynamical ensembles, but in fact $\alpha$ is related to entropy only in the reversible model, whereas it bears no connection with it in the irreversible one.
Our results confirm this picture with $\alpha$  fluctuating little in the irreversible model and not around $\nu$, as found for the reversible model.

\paragraph{Conclusions} 
We have shown through high-resolved numerical simulations that the Gallavotti's conjecture of dynamical ensemble equivalence is correct.
We observe that no matter the Reynolds number, provided sufficient resolution is kept, not only the basic requirements of the conjecture are fulfilled, but all the  one- and two-point statistical observables are found indistinguishable in the irreversible and reversible dynamical system.
Furthermore, the scale-by-scale analysis of the kinetic energy flux again shows negligible difference between the two models up to the dissipation range, far beyond the original formal conjecture proposition.
Wild fluctuations of the fluctuating viscosity are encountered and at high-Re numbers, even negative values are recorded, which point out to local anti-dissipative phenomena. However, these negative events remain extremely rare.
Our results confirm preliminary results obtained in simplified dynamical models of turbulence \cite{biferale2018equivalence}.

Our results give empirical evidence that the \emph{chaotic hypothesis} from which the conjecture is originally derived can be considered \emph{morally} applicable to turbulent fluids. That means in turn that non-equilibrium statistical mechanics and notably fluctuation relations about entropy production should apply in some sense also to turbulent fluids.
Our results corroborate the dynamical system approach to turbulence as main theoretical framework for turbulence \cite{ruelle1995turbulence,ruelle2012hydrodynamic}.

Furthermore, it is shown that turbulence is unaffected by the precise mechanism of dissipation.
This is a conceptually important result, since it corroborates 
the idea that scales larger than the forcing are governed by Euler, as recently proposed~\cite{dallas2015statistical,michel2017observation,dallas2020transitions}.
On the other hand, it paves the way to the use of whatever phenomenological model, provided the correct amount of average rate of dissipation is enforced.

Some issues remain to be answered, while the reversible system appears mathematically simpler because of the constraint  on the enstrophy, the presence of negative events in viscosity makes it not well-posed, shifting but not solving the question of global existence of the solution. Rigorous analysis lacks.
The possibility to compute non-equilibrium entropy and its behaviour is appealing but the needed statistics to make predictions seems  overwhelming in 3D.
More notably, to exploit the new framework to get new insights on turbulence problem remains an unexplored route.

\paragraph{Acknowledgements}
The authors thank the several deep and fruitful discussions with Giovanni Gallavotti at the early stage of the work. 
This work was granted access to the HPC resources of [TGCC/CINES/IDRIS] under the allocation 2019- [A0062B10759] and 2020- [A0082B10759] attributed by GENCI (Grand Equipement National de Calcul Intensif).

\appendix

\section{Scale-by-scale analysis}
We recall that The Navier-Stokes equations for the velocity field $\bf u$ of an incompressible unit density fluid are given by
\begin{eqnarray}
\ {\partial {\bf u}\over\partial t}+{\bf u}\cdot\nabla{\bf u} &=&
   -\nabla p + 
{\nu}\nabla^2\ {\bf u +f} \\
\nabla &\cdot& {\bf u}=0,
  \label{momentum}
\end{eqnarray}
where $p$ is the pressure, $\nu$ is the viscosity and $\bf f$ an external body force.

We introduce the notion of different scales in the flow using the filtering or coarse-graining approach \cite{germano1992turbulence}, where the dynamic velocity field $\bu$ is spatially (low-pass) filtered over a scale $\ell$ to obtain 
the filtered velocity field $\fltr{\bu}_\ell(\bx)$. 
The filtering procedure is given by  
\be 
\fltr{\bu}_\ell(\bx) = \int d^3 r\, G_\ell(\br) \bu(\bx+\br) 
\label{eq:filter}
\ee
where $G_\ell$ is a smooth filtering function, spatially localized and such that $G_\ell (\vec r) = \ell^{-3}G(\vec r/\ell)$ where the function $G$ satisfies
$\int d\vec r \ G(\vec r)=1$, and $\int d\vec r \ \vert \vec r \vert ^2 G(\vec r) = \mathcal{O}(1)$. 
By applying such a coarse-graining to the Navier-Stokes equations we obtain:
\be 
\partial_{t} \widetilde{\bu}_\ell 
+  (\widetilde{\bu}_\ell \cdot \grad)\widetilde{\bu}_\ell = -\grad\widetilde{p}_{\ell} 
-\grad\cdot\btau_\ell
+\nu\nabla^{2}\widetilde{\bu}_\ell.
\label{eq:u-eq-ell} 
\ee
This equation describes the dynamics at the scale $\ell$, and 
  $\btau_\ell$ is the 
{\it subscale stress-tensor} (or momentum flux) which describes the force exerted 
on scales larger than $\ell$ by fluctuations at scales smaller than $\ell$. It is given by:
\be 
(\btau_\ell)_{i,j} = 
\widetilde{(u_i u_j)}_\ell -
(\widetilde{u}_\ell)_i (\widetilde{u}_\ell)_j 
\ee 
The corresponding pointwise kinetic energy budget reads
\begin{eqnarray}
\partial_t \left(\frac{1}{2}|\fu_\ell|^2\right) &+& \partial_j\left[ \left(\frac{1}{2}|\fu_\ell|^2 +\widetilde{p}_\ell \right)) (\widetilde{u}_\ell)_j  + \tau_{ij}(\widetilde{u}_\ell)_i
- \nu\partial_j\left(\frac{1}{2}|\fu_\ell|^{2}\right)\right] \\ \nonumber
&=& -\Pi_\ell  
 - \nu|\grad\fu_\ell|^2,
\lb{kinetic-large}
\end{eqnarray}
 with 
\be \Pi_\ell(\bx) \equiv -(\partial_{j}\widetilde{u}_{i})\tau_{ij}. 
\lb{kinetic-flux}
\ee
 the sub-grid scale (SGS) energy flux. This term is key since it represents the space-local transfer of energy among large and small scales across the scale $\ell$. In the case of direct energy cascade, the flux is known to be positive in average.

The present scale-by-scale procedure holds in the physical space, however an efficient way to implement the filter in homogeneous flows is through the Fourier transform 
\be \hat{G}_q (\bk) = \int G_\ell({\bf x}) e^{i\bf k\cdot x}  d {\bx} \ee
where $q=1/\ell$ is the filtering wavenumber. 
In this work we have considered a Gaussian kernel 
\be \hat{G}_q(\bk)=\exp\left[-\frac{k^2}{2q^2}\right]. 
\label{gauss}\ee 
For an infinite domain this filter corresponds to the Gaussian filter in real space $G_\ell(r) =\exp(-\frac{1}{2}r^2/\ell^2) /(2\pi \ell^{2})^{3/2}$ %\NOTE{I need to Check $2\pi$ factors!}. 

\section{Equivalence of dynamical ensembles}
We give here a brief account of the foundation of the Gallavotti's conjecture.
The presentation closely follows the original developments \cite{gallavotti2013foundations,gallavotti2014nonequilibrium,gallavotti2019nonequilibrium}, to which we refer for a better description.

Let us consider a smooth dynamical system on a phase-space $\mathcal M$, which may depend on several parameters $\mathbf{P}$.
We consider that the system generates a Sinai-Ruelle-Bowen (SRB) stationary state \cite{sinai1968markov,bowen1975ergodic} for each value of the parameters,
and the collection $\mathcal E$ of such probability distributions $\mu_\mathbf{P}$ constitutes an ensemble.
In general, more than a single SRB distribution (which would mean more than an attractor) can be generated for a given set of parameters, but we do not consider this possibility for the sake of clarity, without loss of generality.
Empirically, we have carried out several tests and we have not found any case for which different attractors are met in our numerical set up.

It is supposed that the following hypothesis holds\cite{gallavotti1995dynamical,gallavotti2019nonequilibrium}:

\emph{Chaotic Hypothesis (CH): A chaotic evolution takes place on a phase-space $\mathcal M$ being attracted by a bounded smooth attracting surface $\mathcal A\subset M$ and on $\mathcal A$ the flow is Anosov.}

The conceptual content of this hypothesis is that all the systems sufficiently chaotic (in the Lyapunov sense) can be treated ``in practice" as Anosov systems. In some sense, that is analogous to the ergodic hypothesis for the equilibrium statistical mechanics, which is not rigorously true but can be considered as true for macroscopic bodies \cite{lifshitz2013statistical,castiglione2008chaos,chibbaro2014reductionism}.

In the present work, we have considered the Navier-Stokes equations which has only one parameter: the viscosity $\mathbf{P}=\{\nu\}$; and thus we have a corresponding SRB distribution $\mu_\nu$.
 We have yet considered that it is possible to propose a different model parametrized by another parameter $\Omega$, which gives a description \emph{equivalent} to the Navier-Stokes model.
The equivalence means here that both models
give the same predictions for a large class of relevant observables. 
 The alternative model generates a new ensemble of SRB distributions $\mu_\Omega \in \mathcal E^\prime$.
 In this sense, we may have equivalent dynamical nonequilibrium ensembles.
 
 This proposition is analogous to (and a generalisation of)  the Gibbs ensemble description of statistical mechanics of equilibrium systems. In particular, there is the microcanonical ensemble  given by the distribution $\mu_E$, which depends on the fixed energy $E$, and the canonical distribution $\mu_\beta$ depending on the fixed (inverse) temperature $\beta=(k_B T)^{-1}$.
 These ensembles are equivalent in the sense that in the thermodynamic limit, where the number of molecules $N \rightarrow \infty$, the average of most of the observables $\mathcal O$ is the same in both ensembles $\mu_E(\mathcal O)=\mu_\beta(\mathcal O)$.
 
 In the main text, we have proposed an equivalent model, for which the parameter is the enstrophy $\Omega$.
 In this model, the enstrophy is a constant of motion, but to fulfil such constraint the constant fluid viscosity $\nu$ is replaced by a fluctuating multiplier $\alpha$.
 In our case, the analogous of the thermodynamic limit is considered as usual in turbulence to be the large Reynolds or vanishing viscosity limit, and therefore the equivalence is expressed as
 \begin{equation}
\lim_{\nu \rightarrow 0} \mu_\nu(\mathcal O) = \lim_{\nu \rightarrow 0} \mu_\Omega(\mathcal O)~,
\end{equation}
provided that $\nu=\mu_\Omega(\alpha)$.

\bibliographystyle{abbrv}
\bibliography{biblio}
%----------------------------------------------------------------------------
%----------------------------------------------------------------------------

\end{document}